\documentclass[doublecol]{epl2}

\usepackage{graphicx}  
\usepackage{amsmath}
\usepackage{amssymb}
\usepackage{multirow}

\usepackage[numbers,sort&compress,mcite]{natbib}

\title{Optimize quantum simulation using a force-gradient integrator}
\shorttitle{Optimize quantum simulation using a force-gradient integrator} 

\author{Yi-Tong Zou\inst{1} \and Yu-Jiao Bo\inst{1} \and Ji-Chong Yang\inst{1}}
\shortauthor{Y.-T. Zou \etal}

\institute{
  \inst{1} Department of Physics, Liaoning Normal University - Dalian 116029, China
}

\abstract{
Quantum simulation has shown great potential in many fields due to its powerful computational capabilities.
However, the limited fidelity can lead to a severe limitation on the number of gate operations, which requires us to find optimized algorithms.
Trotter decomposition and high order Trotter decompositions are widely used in quantum simulations.
We find that they can be significantly improved by force-gradient integrator in lattice QCD.
By using two applications as examples, we show that the force-gradient decomposition can reduce the number of gate operations up to about a third of those using high order Trotter decompositions.
Therefore, force-gradient decomposition shows a great prospective in future applications of quantum simulation.
}

\begin{document}



\maketitle


\section{\label{level1}Introduction}

Quantum simulation as proposed by Feynman~\cite{feynman1981}, has shown great potential in many fields due to its powerful computational capabilities and their potential to avoid the fermionic sign problem~\cite{qsreview1,*sign1,*sign2} which is an NP hard problem~\cite{signNPhard}.
Consequently, in recent years quantum simulation has developed rapidly and been applied in many fields~\cite{simulation1,*simulation2,*simulation3,*simulation4}.
With Google's announcement of quantum `supremacy'~\cite{google} and the subsequent experiment using photons~\cite{Jiuzhang}, quantum simulations hold great promise for the future.

One of the biggest problems plaguing quantum simulations is the lack of fidelity.
Recently, the fidelity can reach up to $99.9\%$ for one-qubit gate and $94.7\%$ for Clifford gate~\cite{fidelity}.
The fidelity decays exponentially with the number of gate operations, as a result the required computing resources increase exponentially, and the computational power of a quantum computer will be unable to surpass that of classical computers~\cite{PhysRevX.10.041038}.
This problem has led to widespread researches in quantum error correction~\cite{qec1,*qec2,*qec3,*qec5,*qec7,*qec8,*qec9,*qec10}.

The Trotter~(or Lie–Trotter–Suzuki) decomposition~(TD) is one of the earliest quantum algorithms in quantum simulations~\cite{Lloyd1996}.
TD belongs to the product formulas, although some post-Trotter methods have been proposed, the product formulas are still found to be competitive especially in the case that the simulated system has a Lie-algebraic structure, and the error of product formulas has been well studied~\cite{trottererror}.
A class of optimizations of TD is the high-order TDs~\cite{bookhighorder1984,*Suzuki1991}, for example the seconder order~(or symmetric) TD~(STD) has been widely applied~\cite{std1,*std2,*std3,*std4,std5}.
Until very recently, TD was still used in simulations~\cite{timtd1,*td3,*timtd2,*d4}.
The optimization of TD and high-order TDs is very important, since the fidelity decays exponentially, even modest optimization can make a drastic improvement, and may enable some models to be simulated under existing conditions.
In addition, the real quantum computers are not yet widespread and many simulations are performed on quantum computer simulators running on a classical computer~\cite{classicalsimulation1}.
The optimization can accelerate these simulations significantly as well.

There is a problem in the field of lattice QCD similar to the improvement on TD and high order TDs.
The STD with two non-commutating terms can be corresponded to the leapfrog integrator in lattice QCD.
However, there are integrators much faster than leapfrog such as Omelyan integrator~\cite{omelyan} and force-gradient integrators~\cite{fg1,fg2,fg3}.
The latter does not belong to the class of high-order TDs.
There has been prior works on optimizing the simulations of quantum systems using analytically-derived gradients, such as the GRAPE algorithm~\cite{gradientascent1,gradientascent4,gradientascent2,gradientascent3}.
However, to our knowledge the force-gradient integrators have barely caught the attention of the quantum simulation community so far.
The force-gradient integrators can be applied in quantum simulations alongside other optimizations used in product formulas~\cite{std5,optimization1,*optimization2,*optimization3}, and with the nested integrator techniques~\cite{nested}.
In this letter, we investigate the feasibility of a force-gradient decomposition~(FGD) in the quantum simulations.

\section{\label{level2}Force-gradient decomposition}

Considering a system whose Hamiltonian can be written as $H=S+T$ with $\left[ S,T \right] \neq 0$, such a system can be simulated using TD, i.e. the $\exp{\left({\rm i}Ht\right)}$ is approximated by $\exp ({\rm i}Ht) \approx(\exp ({\rm i} \tau S) \exp ({\rm i} \tau T))^{m}$ with $\tau=t/m$.
When decomposed to $m$ steps, the total number of exponential operations~(denoted as $n$) required is $n=2m$.
Its error can be roughly estimated by the Baker-Campbell-Hausdorff formula.
Assume $\tau < 1$ and $\left[ S,T \right] \sim \mathcal{O}(l)$, the error when evolute to $t$ is $\epsilon \sim \mathcal{O}\left( t^2 l /2m \right)$.
Similarly, the STD is $\exp ({\rm i}Ht) \approx \left(\exp \left({\rm i}\tau S/2\right) \exp ({\rm i}\tau T) \exp \left({\rm i}\tau S/2\right)\right)^m$ with $n=2m+1$ and $\epsilon \sim \mathcal{O}\left( t^3 l^2 /8m^2 \right)$.
The STD is in the category of higher-order TDs which has the form $ {\rm e}^{t_1S}{\rm e}^{t_2T}{\rm e}^{t_3S}{\rm e}^{t_4T}\ldots {\rm e}^{t_MS}$~\cite{bookhighorder1984,*Suzuki1991}, the FGDs are no longer higher-order TDs. A widely used FGD is~\cite{fg3}
\begin{equation}
\begin{split}
&\exp \left\{{\rm i}m \tau\left(S+T+\mathcal{O}\left(\tau^{4}\right)+\mathcal{O}\left(\tau^{6}\right)\right)\right\}=\\
&\left\{\exp \left(\frac{{\rm i}\tau S}{6}\right) \exp \left(\frac{{\rm i}\tau T}{2}\right) \exp \left(\frac{2{\rm i} \tau S}{3}+\frac{{\rm i}\tau^{3}}{72}[S,[S, T]]\right) \right.\\
&\times \left.\exp \left(\frac{{\rm i}\tau T}{2}\right) \exp \left(\frac{{\rm i}\tau S}{6}\right)\right\}^{m},
\end{split}
\label{eq.2.1}
\end{equation}
with
\begin{equation}
\begin{split}
&\mathcal{O}\left(\tau^{4}\right)=-\frac{\tau^{4}}{155520}\left\{41\left[~S,[~S,[~S,[~S, ~T]]]\right]\right.\\
&\left.+36[[~S, ~T],[S[S, T]]]+72[[~S, ~T],[T,[S, T]]]\right.\\
&\left.+84[~T,[~S,[~S,[~S, ~T]]]]+126[~T,[~T,[~S,[~S, ~T]]]]\right.\\ &\left.+54[~T,[~T,[~T,[~S, ~T]]]]\right\}.
\end{split}
\label{eq.2.2}
\end{equation}
Usually, $[S, [S,[S,T]]]\neq 0$, therefore one needs to further decompose $\exp \left(2{\rm i}\tau S/3 +{\rm i} \tau^3 [S,[S,T]]/72 \right)$.
Both $n$ and $\epsilon$ depend on how this term is decomposed.
Nevertheless, as will be shown, Eq.~(\ref{eq.2.1}) can be usually decomposed as the $m$-th power of the product of $7$ exponents~(denoted as 7-stage decomposition).
For $l$-stage decomposition, $n=(l-1)m$.

A general discussion of the error of FGD is beyond our ability, and is model dependent.
Instead, we study the optimization brought by the FGD using two specific applications as examples.

\section{\label{level3}Applications}

To concentrate on the error due to the decomposition, we define $\varepsilon= \sqrt{||\exp({\rm i}tH)-M||}/\sqrt{||\exp({\rm i}tH)||}$, where $||\ldots||$ denotes the sum of squares of the matrix elements and the matrix $M$ denotes the decomposed matrix, for example $M=\left(\exp{\left({\rm i}\tau S\right)}\exp{\left({\rm i}\tau T\right)}\right)^m$ in the case of TD.

For a small system, the Hamiltonian can be numerically diagonalized, and $\varepsilon$ can be calculated on a classical computer.
We use two small systems to compare different decompositions.
The Omelyan integrator is frequently used in lattice QCD and is also included as an example of high-order TDs, which is denoted as Omelyan decomposition~(OD)~\cite{omelyan},
\begin{equation}
\begin{split}
\exp ({\rm i}tH) &\approx\left\{\exp \left({\rm i} \alpha \tau S\right) \exp \left(\frac{{\rm i}\tau T}{2}\right) \right.\\
&\left.\times \exp \left({\rm i} (1-2\alpha)\tau S\right)\exp \left(\frac{{\rm i}\tau T}{2}\right)\exp \left({\rm i} \alpha \tau S\right)\right\}^{m},
\end{split}
\label{eq.3.1}
\end{equation}
with $\alpha \approx 0.1931833275037836$.

The FGD in Eq.~(\ref{eq.2.1}) is usually a 7-stage decomposition, therefore we also compare FGD with a 7-stage high order Trotter decomposition~(7TD).
An optimized 7TD is~\cite{fg2}
\begin{equation}
\begin{split}
\exp \left({\rm i}tH\right)&\approx
\left\{\exp \left({\rm i}\frac{\beta}{2} \tau S\right) \exp \left({\rm i}\beta \tau T\right) \exp \left({\rm i}\frac{1-\beta}{2}\tau S\right)\right.\\
&\left.\times \exp \left({\rm i}(1-2\beta) \tau T\right) \exp \left({\rm i}\frac{1-\beta}{2}\tau S\right)\right.\\
&\left.\times \exp \left({\rm i}\beta \tau T\right) \exp \left({\rm i}\frac{\beta}{2} \tau S\right)\right\}^{m},\\
\beta &=\frac{1}{2-\sqrt[3]{2}}
\end{split}
\label{eq.3.2}
\end{equation}

Since the number of gate operations is approximately proportional to $n$ for a same model, we use $n$ to quantify the complexity of the decompositions.

\subsection{\label{level3.1}Transverse Ising model}

The transverse Ising model~(TIM) is a popular benchmark for quantum simulations.
Therefore we also use TIM in $1D$ and $2D$ to test the FGD.

The Hamiltonian of TIM can be written as
\begin{equation}
H=\sum _{\langle ij\rangle} \sigma_{z}\left(n_i\right) \sigma_{z}\left(n_j\right)+\lambda \sum_i \sigma_{x}\left(n_i\right),
\label{eq.3.3}
\end{equation}
where $\sigma_{x,z}$ are the Pauli matrices, $\langle ij\rangle$ refers to the nearest neighbouring pairs.
The $\sigma _z(n_i)$ is short for the tensor product of $2\times 2$ matrices on each site, the matrix on $n_i$ is $\sigma _z$, and the others are identity matrices.

Firstly, we consider a $1D$ transverse Ising chain with only $3$ sites and with a periodic boundary condition.
The Hamiltonian in this case is a $8\times 8$ matrix acting on $3$ qubits.
Denoting the indices of the sites as $n_{1,2,3}$, $H=T+S$ with
\begin{equation}
\begin{split}
&T=\sigma_{z}\left(n_{1}\right) \sigma_{z}\left(n_{2}\right)+\sigma_{z}\left(n_{2}\right) \sigma_{z}\left(n_{3}\right)+\sigma_{z}\left(n_{3}\right) \sigma_{z}\left(n_{1}\right),\\
&S=\lambda \sum_{i=1}^3 \sigma_{x}\left(n_{i}\right).
\end{split}
\label{eq.3.4}
\end{equation}
Therefore $[S,[S, T]]=-8 \lambda^{2}(Y-T)$ with
\begin{equation}
Y=\sigma_{y}\left(n_{1}\right) \sigma_{y}\left(n_{2}\right)+\sigma_{y}\left(n_{2}\right) \sigma_{y}\left(n_{3}\right)+\sigma_{y}\left(n_{3}\right) \sigma_{y}\left(n_{1}\right).
\label{eq.3.5}
\end{equation}
For the $\exp \left(2{\rm i}\tau S/3 +{\rm i} \tau^3 [S,[S,T]]/72 \right)$ term we use the STD which is found to be optimized among 5-stage decompositions for $3$ non-commutating terms~\cite{threeterm}.
The FGD is then
\begin{equation}
\begin{split}
\exp ({\rm i}tH) &\approx\left\{\exp \left(\frac{{\rm i} \tau S}{6}\right) \exp \left({\rm i} T\left(\frac{\tau}{2}+\frac{\tau^{3} \lambda^{2}}{18}\right)\right) \right.\\
&\left.\times \exp \left(\frac{{\rm i} \tau S}{3}\right)\exp \left(-{\rm i} \frac{\tau^{3} \lambda^{2}}{9} Y\right) \exp \left(\frac{{\rm i} \tau S}{3}\right) \right.\\
&\left.\times \exp \left({\rm i} T\left(\frac{\tau}{2}+\frac{\tau^{3} \lambda^{2}}{18}\right)\right) \exp \left(\frac{{\rm i} \tau S}{6}\right)\right\}^{m}.
\end{split}
\label{eq.3.6}
\end{equation}
Similar to the TD, before each exponential evaluation, the qubits should be rotated to $\sigma _{x,y,z}$ representations accordingly. Note that there is a $Y$ operator which is diagonalized in $\sigma _y$ representation which is different from TD and high order TDs.

\begin{table}[!htbp]
\begin{center}
\begin{tabular}{|c|c|c|c|c|c|c|}
\hline
 $\lambda$ & & TD & STD & OD & 7TD & FGD \\
 \hline
 \multirow{2}*{$0.5$}
~ & $n_{\rm min}$ & $1034$ & $39$ & $33$ & $43$ & $19$ \\
\cline{2-7}
~ & $\varepsilon(\%)$ & $0.10$ & $0.098$ & $0.086$ & $0.061$ & $0.033$ \\
\hline
\multirow{2}*{$1$}
~ & $n_{\rm min}$ & $1606$ & $55$ & $53$ & $55$ & $25$ \\
\cline{2-7}
~ & $\varepsilon(\%)$ & $0.10$ & $0.098$ & $0.099$ & $0.088$ & $0.035$ \\
\hline
\multirow{2}*{$1.5$}
~ & $n_{\rm min}$ & $1456$ & $71$ & $69$ & $67$ & $31$ \\
\cline{2-7}
~ & $\varepsilon(\%)$ & $0.10$ & $0.095$ & $0.094$ & $0.092$ & $0.048$ \\
\hline
\end{tabular}
\end{center}
\caption{\label{Tab:IsingChain}The $n_{\rm min}$ required to satisfy $\varepsilon < 0.1\%$ for the Ising chain when $t=1$.}
\end{table}

In the case of infinite volume, the $1D$ TIM is self dual, there is a phase transition at $\lambda _c=1$~\cite{KogutReview,TIMBook}.
Therefore we use $\lambda = 0.5, 1, 1.5$ as examples.
When $t=1$, we calculate the minimum number of exponential operations needed~(denoted as $n_{\rm min}$) when the error satisfies $\varepsilon < 0.1\%$.
The results are shown in Table~\ref{Tab:IsingChain}.

\begin{figure}[!htbp]
\centering
\includegraphics[width=0.48\textwidth]{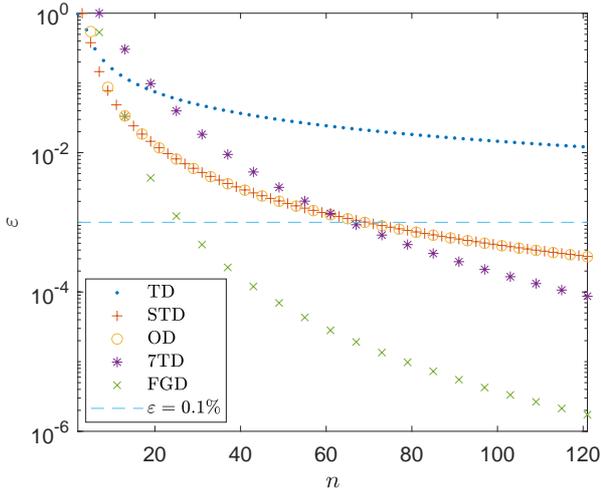}
\caption{\label{fig.errorIsing}$\varepsilon$ as functions of $n$ for Ising chain when $t=1,\lambda = 1.5$.}
\end{figure}
The non-commutativity $l$ grows with $\lambda$, therefore the worst case is when $\lambda =1.5$.
When $\lambda =1.5,t=1$, the error decays for the decompositions are shown in Fig.~\ref{fig.errorIsing}.
The results indicates that the simulation of $1D$ TIM can be significantly optimized by the FGD.

Once the FGD of $1D$ TIM is known, the FGD for higher dimensional TIM with periodic boundary condition can be obtained because $T=\sum T_{1D}$, and $[S,[S,T]]=\sum [S,[S, T_{1D}]]$.

\begin{figure}[!htbp]
\centering
\includegraphics[width=0.3\textwidth]{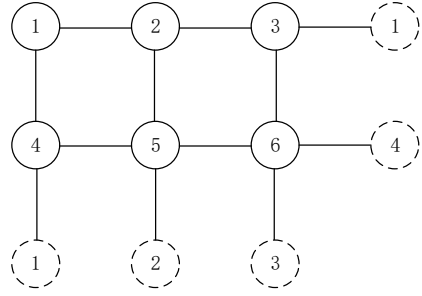}
\caption{\label{fig.figd2ising}The TIM on a $2\times 3$ lattice with periodic boundary condition, the numbers are indices of the sites.}
\end{figure}

\begin{figure}[!htbp]
\centering
\includegraphics[width=0.48\textwidth]{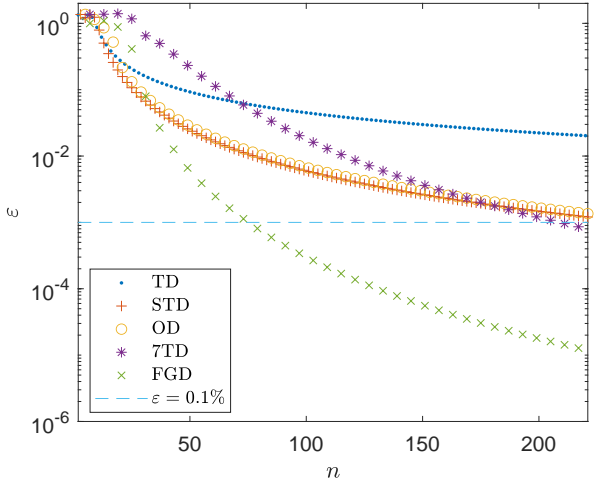}
\caption{\label{fig.errorIsing2d}$\varepsilon$ as functions of $n$ for $2D$ TIM when $t=1,\lambda = 5$.}
\end{figure}

Take the TIM on a $2\times 3$ lattice with periodic boundary condition as an example, the lattice is depicted in Fig.~\ref{fig.figd2ising}, and the Hamiltonian is a $64\times 64$ matrix with
\begin{equation}
\begin{split}
&S=\sum _{i=1}^6 \sigma _x (n_i),\;\;T=\sum _{j=1}^5 T_j,\;\;Y=\sum _{j=1}^5Y_j,\\
&T_1=\sigma_z (n_1)\sigma_z (n_2)+\sigma_z (n_2)\sigma_z (n_3)+\sigma_z (n_3)\sigma_z (n_1),\\
&T_2=\sigma_z (n_4)\sigma_z (n_5)+\sigma_z (n_5)\sigma_z (n_6)+\sigma_z (n_6)\sigma_z (n_4),\\
&T_3=2\sigma_z (n_1)\sigma_z (n_4),\;\;
 T_4=2\sigma_z (n_2)\sigma_z (n_5),\\
&T_5=2\sigma_z (n_3)\sigma_z (n_6),\\
&Y_1=\sigma_y (n_1)\sigma_y (n_2)+\sigma_y (n_2)\sigma_y (n_3)+\sigma_y (n_3)\sigma_y (n_1),\\
&Y_2=\sigma_y (n_4)\sigma_y (n_5)+\sigma_y (n_5)\sigma_y (n_6)+\sigma_y (n_6)\sigma_y (n_4),\\
&Y_3=\sigma_y (n_1)\sigma_y (n_4),\;\;
 Y_4=\sigma_y (n_2)\sigma_y (n_5),\\
&Y_5=\sigma_y (n_3)\sigma_y (n_6).\\
\end{split}
\label{eq.3.7}
\end{equation}
It can be verified that $[S,[S,T]]=-8\lambda ^2 (Y-T)$ holds.
Therefore one can still use the FGD in Eq.~(\ref{eq.3.6}).
The critical transverse field $\lambda _c$ for $2D$ TIM is found to be $2\sim 4$~\cite{TIMBook,timkc1,*timkc2,*timkc3,*timkc4}, therefore we choose $\lambda = 1.5, 3, 5$ as examples.
When $t=1$, $n_{\rm min}$ required when the error satisfies $\varepsilon < 0.1\%$ are listed in Table~\ref{Tab:Ising2D}.
When $\lambda =5,t=1$, the error decays for the decompositions are shown in Fig.~\ref{fig.errorIsing2d}.
Similar as the case of $1D$ TIM, the FGD can significantly reduce the number of gate operations.

\begin{table}[!htbp]
\begin{center}
\begin{tabular}{|c|c|c|c|c|c|c|}
\hline
 $\lambda$ & & TD & STD & OD & 7TD & FGD \\
 \hline
 \multirow{2}*{$1.5$}
~ & $n_{\rm min}$ & $3788$ & $121$ & $125$ & $109$ & $37$ \\
\cline{2-7}
~ & $\varepsilon(\%)$ & $0.10$ & $0.10$ & $0.094$ & $0.088$ & $0.099$ \\
\hline
\multirow{2}*{$3$}
~ & $n_{\rm min}$ & $4042$ & $187$ & $197$ & $157$ & $55$ \\
\cline{2-7}
~ & $\varepsilon(\%)$ & $0.10$ & $0.098$ & $0.098$ & $0.089$ & $0.091$ \\
\hline
\multirow{2}*{$5$}
~ & $n_{\rm min}$ & $4564$ & $243$ & $257$ & $211$ & $79$ \\
\cline{2-7}
~ & $\varepsilon(\%)$ & $0.099$ & $0.10$ & $0.10$ & $0.096$ & $0.081$ \\
\hline
\end{tabular}
\end{center}
\caption{\label{Tab:Ising2D}The $n_{\rm min}$ required to satisfy $\varepsilon < 0.1\%$ for the Ising chain when $t=1$.}
\end{table}

\subsection{\label{level3.2}Ising gauge model}

\begin{figure}[!htbp]
\centering
\includegraphics[width=0.3\textwidth]{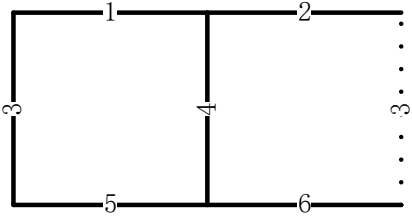}
\caption{\label{fig.z2lattice}The Ising gauge model with two plaquettes, the numbers are indices of the links.}
\end{figure}
The $D=d+1$ dimensional Ising gauge model, i.e. $\mathbb{Z}_2$ lattice gauge model~\cite{z2lgt} at finite temperature can be dual to a $d$-dimension Ising gauge model in a transverse field at zero temperature~\cite{KogutReview,Fradkin1978}, also known as one of the quantum link models~\cite{quantumlinkmodel}.
The Hamiltonian can be written as
\begin{equation}
H=\sum _{\square} \prod _{l\in \square}\sigma_{z}\left(l\right) +k \sum_{l} \sigma_{x}\left(l\right),
\label{eq.3.8}
\end{equation}
where $l$ is an index of a link, and `$\square$' denotes a plaquette.
We consider the case of a stripe containing two plaquettes with periodic boundary condition on major direction as shown in Fig.~\ref{fig.z2lattice}.
The Hamiltonian is a $64\times 64$ matrix with
\begin{equation}
\begin{split}
&S=\prod _{i=1,4,5,3} \sigma _z(l_i)+\prod _{i=2,3,6,4} \sigma _z(l_i),\\
&T=\lambda \sum_{i=1}^6 \sigma_{x}\left(l_{i}\right),\;\;[S,[S,T]]=A+B\\
&A=4\lambda \left(\sum _{i=1,2,5,6}\sigma_{x}\left(l_{i}\right)+2\sigma_{x}\left(l_{3}\right)+2\sigma_{x}\left(l_{4}\right)\right),\\
&B=8\lambda \left(\prod _{i=1,2,5,6} \sigma _z(l_i)\right)\left(\sigma_{x}\left(l_{3}\right)+\sigma_{x}\left(l_{4}\right)\right).
\end{split}
\label{eq.3.9}
\end{equation}
The FGD is
\begin{equation}
\begin{split}
\exp ({\rm i}tH) &\approx\left\{\exp \left(\frac{{\rm i} \tau S}{6}\right) \exp \left(\frac{{\rm i}\tau}{2}T+\frac{{\rm i}\tau^3 A}{144}\right) \right.\\
&\left.\times \exp \left(\frac{{\rm i} \tau S}{3}\right) \exp \left({\rm i} \frac{\tau^{3} \lambda}{72} B\right) \exp \left(\frac{{\rm i} \tau S}{3}\right)\right.\\
&\left.\times \exp \left(\frac{{\rm i}\tau}{2}T+\frac{{\rm i}\tau^3 A}{144}\right) \exp \left(\frac{{\rm i} \tau S}{6}\right)\right\}^{m}.
\end{split}
\label{eq.3.10}
\end{equation}

\begin{table}[!htbp]
\begin{center}
\begin{tabular}{|c|c|c|c|c|c|c|}
\hline
 $k$ & & TD & STD & OD & 7TD & FGD \\
 \hline
\multirow{2}*{$0.1$}
~ & $n_{\rm min}$ & $376$ & $15$ & $13$ & $19$ & $13$ \\
\cline{2-7}
~ & $\varepsilon(\%)$ & $0.10$ & $0.092$ & $0.074$ & $0.035$ & $0.036$ \\
\hline
\multirow{2}*{$0.3$}
~ & $n_{\rm min}$ & $1012$ & $29$ & $25$ & $31$ & $19$ \\
\cline{2-7}
~ & $\varepsilon(\%)$ & $0.10$ & $0.094$ & $0.087$ & $0.067$ & $0.022$ \\
\hline
\multirow{2}*{$1.0$}
~ & $n_{\rm min}$ & $1590$ & $63$ & $53$ & $67$ & $25$ \\
\cline{2-7}
~ & $\varepsilon(\%)$ & $0.10$ & $0.094$ & $0.10$ & $0.090$ & $0.10$ \\
\hline
\end{tabular}
\end{center}
\caption{\label{Tab:z2}The $n_{\rm min}$ required to satisfy $\varepsilon < 0.1\%$ for the Ising gauge model when $t=1$.}
\end{table}

\begin{figure}[!htbp]
\centering
\includegraphics[width=0.48\textwidth]{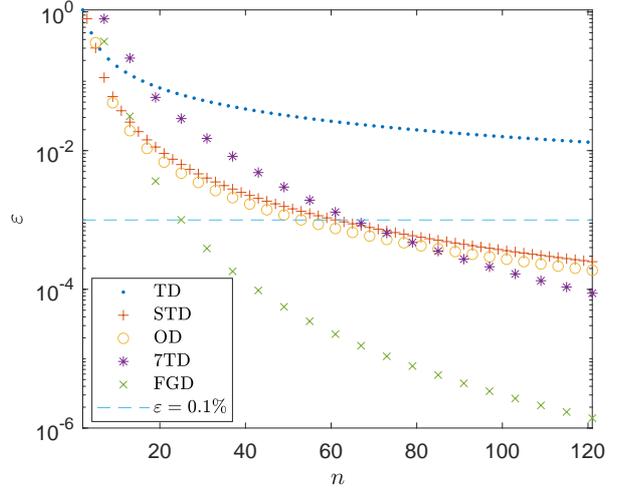}
\caption{\label{fig.errorz2}$\varepsilon$ as functions of $n$ for Ising gauge model when $t=1, k=1$.}
\end{figure}

In the infinite volume, the $2D$ Ising gauge model in transverse field can be dual to $2D$ TIM or $D=2+1$ Ising model at finite temperature~\cite{KogutReview,z2dualtoTIM}.
The $1/k$ can be correspond to $\lambda$ in Eq.(\ref{eq.3.3}).
The $\lambda _c$ for $2D$ TIM is $2\sim 4$, indicating a phase transition at $k_c= 0.25\sim 0.5$, which is also a topological phase transition~\cite{z2Sachdev}.
Recently, $2D$ Ising gauge model is studied by quantum simulation on a $3\times 3$ lattice with periodic boundary condition and $k_c$ is found to be $0.380$~\cite{z2quantum}.
Therefore we use $k =0.1,0.3,1$ as examples.
When $t=1$, the $n_{\rm min}$ needed when $\varepsilon < 0.1\%$ are shown in Table~\ref{Tab:z2}.
For the case of the largest $k$, the error decays are shown in Fig.~\ref{fig.errorz2}.
Again, the FGD can significantly reduce the number of gate operations.

\subsection{\label{level3.3}Summary of the applications}

It can be shown that, the FGD is feasible for many models.
For example, the TIM with periodic condition, and the $2D$ Ising gauge model with two plaquttes.
For both cases, the FGD surpasses 7TD, OD, STD and TD.
When $t=1$, the $n_{\rm min}$ of FGD when $\varepsilon < 0.1\%$ can reach up to about one third of those of TD and high order TDs.
In the case of Ising gauge model with $t=0,k=0.1$, although $n_{\rm min}$ is same for OD and FGD, the $\varepsilon$ of FGD is smaller than half of the $\varepsilon$ of OD.
From the decays of $\varepsilon$, one can see that the optimization mentioned above can be even better when $t$ is larger, or when the required $\varepsilon$ is smaller.

\section{\label{level4}Conclusion}

Quantum simulation is an extremely promising research direction extensively studied recently due to the rapid development of quantum computing technology.
Reducing the number of decomposition steps in the product formulas paves a way to the practical quantum simulation.
While the TD and high order TDs are widely applied in quantum simulations, we show that significant optimization can be achieved by the force-gradient integrator used in lattice QCD.
We use the TIM and Ising gauge model as examples, the FGD can reduce the number of gate operations up to about a third of those using high order TDs.
In addition, it can be seen that if one wishes to use FGD, the $\exp \left(2{\rm i}\tau S/3 +{\rm i} \tau^3 [S,[S,T]]/72 \right)$ term needs to be processed, which is sometimes not easy to handle.
Nevertheless, FGD shows great advantages in quantum simulations and deserves further studies for various of models.

\begin{acknowledgments}
This work was partially supported by the National Natural Science Foundation of China under Grant No. 12047570.
\end{acknowledgments}

\bibliography{ForceGradient}

\begin{thebibliography}{10}
\expandafter\ifx\csname url\endcsname\relax\def\url#1{\texttt{#1}}\fi

\bibitem{feynman1981}
\Name{Feynman R.~P.} \REVIEW{Int. J. Theor. Phys.}{21}{1982}{467}.

\bibitem{qsreview1}
\Name{Georgescu I.~M., Ashhab S. \and Nori F.} \REVIEW{Rev. Mod.
  Phys.}{86}{2014}{153}.

\bibitem{sign1}
\Name{Ortiz G., Gubernatis J.~E., Knill E. \and Laflamme R.} \REVIEW{Phys. Rev.
  A}{64}{2001}{022319}.

\bibitem{sign2}
\Name{Lawrence S.} \Book{{Sign Problems in Quantum Field Theory: Classical and
  Quantum Approaches}} Ph.D. thesis Maryland U. (2020).

\bibitem{signNPhard}
\Name{Troyer M. \and Wiese U.-J.} \REVIEW{Phys. Rev. Lett.}{94}{2005}{170201}.

\bibitem{simulation1}
\Name{Yan Z., Zhang Y.-R., Gong M., Wu Y., Zheng Y., Li, S~Wang C., Liang F.,
  Lin J., Xu Y., Guo C., Sun L., Peng C.-Z., Xia K., Deng H., Rong H., You
  J.-Q., Nori F., Fan H., Zhu X. \and Pan J.-W.}
  \REVIEW{Science}{364}{2019}{753}.

\bibitem{simulation2}
\Name{Collaborators*\textdagger{} G. A.~Q.,  \etal}
  \REVIEW{Science}{369}{2020}{1084}.

\bibitem{simulation3}
\Name{King A.~D., Carrasquilla J., Ozfidan I., Raymond J., Andriyash E.,
  Berkley A., Reis M., Lanting T.~M., Harris R., Poulin-Lamarre G., Smirnov
  A.~Y., Rich C., Altomare F., Bunyk P., Whittaker J., Swenson L., Hoskinson
  E., Sato Y., Volkmann M., Ladizinsky E., Johnson M., Hilton J. \and Amin
  M.~H.} \REVIEW{Nature}{560}{2018}{456}.

\bibitem{simulation4}
\Name{Sparrow C., Mart\'{i}n-L\'{o}pez E., Maraviglia N., Neville A., Harrold
  C., Carolan J., Joglekar Y.~N., Hashimoto T., Matsuda N., O’Brien J.~L.,
  Tew D.~P. \and Laing A.} \REVIEW{Nature}{557}{2018}{660}.

\bibitem{google}
\Name{Arute F., Arya K., Babbush R., Bacon D., Bardin J.~C., Barends R., Biswas
  R., Boixo S., Brandao F. G. S.~L., Buell D.~A., Burkett B., Chen Y., Chen Z.,
  Chiaro B., Collins R., Courtney W., Dunsworth A., Farhi E., Foxen B., Fowler
  A., Gidney C., Giustina M., Graff R., Guerin K., Habegger S., Harrigan M.~P.,
  Hartmann M.~J., Ho A., Hoffmann M., Huang T., Humble T.~S., Isakov S.~V.,
  Jeffrey E., Jiang Z., Kafri D., Kechedzhi K., Kelly J., Klimov P.~V., Knysh
  S., Korotkov A., Kostritsa F., Landhuis D., Lindmark M., Lucero E., Lyakh D.,
  Mandrà S., McClean J.~R., McEwen M., Megrant A., Mi X., Michielsen K.,
  Mohseni M., Mutus J., Naaman O., Neeley M., Neill C., Niu M.~Y., Ostby E.,
  Petukhov A., Platt J.~C., Quintana C., Rieffel E.~G., Roushan P., Rubin
  N.~C., Sank D., Satzinger K.~J., Smelyanskiy V., Sung K.~J., Trevithick
  M.~D., Vainsencher A., Villalonga B., White T., Yao Z.~J., Yeh P., Zalcman
  A., Neven H. \and Martinis J.~M.} \REVIEW{Nature}{574}{2019}{505}.

\bibitem{Jiuzhang}
\Name{Zhong H.-S., Wang H., Deng Y.-H., Chen M.-C., Peng L.-C., Luo Y.-H., Qin
  J., Wu D., Ding X., Hu Y., Hu P., Yang X.-Y., Zhang W.-J., Li H., Li Y.,
  Jiang X., Gan L., Yang G., You L., Wang Z., Li L., Liu N.-L., Lu C.-Y. \and
  Pan J.-W.} \REVIEW{Science}{370}{2020}{1460}.

\bibitem{fidelity}
\Name{Huang W., Yang C.~H., Chan K.~W., Tanttu T., Hensen B., Leon R. C.~C.,
  Fogarty M.~A., Hwang J. C.~C., Hudson F.~E., Itoh K.~M., Morello A., Laucht
  A. \and Dzurak A.~S.} \REVIEW{Nature}{569}{2019}{532–}.

\bibitem{PhysRevX.10.041038}
\Name{Zhou Y., Stoudenmire E.~M. \and Waintal X.} \REVIEW{Phys. Rev.
  X}{10}{2020}{041038}.

\bibitem{qec1}
\Name{Grassl M. \and R{\"o}tteler M.} \Book{Quantum Error Correction and Fault
  Tolerant Quantum Computing} (Springer New York, New York, NY) 2012 pp.
  2478--2496.

\bibitem{qec2}
\Name{Steane A.~M.} \REVIEW{Phys. Rev. Lett.}{77}{1996}{793}.

\bibitem{qec3}
\Name{Campagne-Ibarcq P., Eickbusch A., Touzard S., Zalys-Geller E., Frattini
  N.~E., Sivak V.~V., Reinhold P., Puri S., Shankar S. \and Schoelkopf R.~J.}
  \REVIEW{Nature}{584}{2020}{368–}.

\bibitem{qec5}
\Name{Kitaev A.} \REVIEW{Annals of Physics}{303}{2003}{2}.

\bibitem{qec7}
\Name{Linke N.~M., Gutierrez M., Landsman K.~A., Figgatt C., Debnath S., Brown
  K.~R. \and Monroe C.} \REVIEW{Science Advances}{3}{2017}{10}.

\bibitem{qec8}
\Name{Rosenblum S., Reinhold P., Mirrahimi M., Jiang L., Frunzio L. \and
  Schoelkopf R.~J.} \REVIEW{Science}{361}{2018}{266}.

\bibitem{qec9}
\Name{Yao X.-C., Wang T.-X., Chen H.-Z., Gao W.-B., Fowler A.~G., Raussendorf
  R., Chen Z.-B., Liu N.-L., Lu C.-Y., Deng Y.-J., Chen Y.-A. \and Pan J.-W.}
  \REVIEW{Nature}{482}{2012}{489}.

\bibitem{qec10}
\Name{Cho A.} \REVIEW{Science}{369}{2020}{130}.

\bibitem{Lloyd1996}
\Name{Lloyd S.} \REVIEW{Science}{273}{1996}{1073}.

\bibitem{trottererror}
\Name{Childs A.~M., Su Y., Tran M.~C., Wiebe N. \and Zhu S.} \REVIEW{Phys. Rev.
  X}{11}{2021}{011020}.

\bibitem{bookhighorder1984}
\Name{John Day~Dollard C. N.~F.} \Book{Product Integration with Application to
  Differential Equations} Encyclopedia of Mathematics and its Applications 10
  (Cambridge University Press) 1984.

\bibitem{Suzuki1991}
\Name{Suzuki M.} \REVIEW{Journal of Mathematical Physics}{32}{1991}{400}.

\bibitem{std1}
\Name{Suzuki M.} \REVIEW{Journal of Mathematical Physics}{26}{1985}{601}.

\bibitem{std2}
\Name{Wecker D., Bauer B., Clark B.~K., Hastings M.~B. \and Troyer M.}
  \REVIEW{Phys. Rev. A}{90}{2014}{022305}.

\bibitem{std3}
\Name{Zohar E., Farace A., Reznik B. \and Cirac J.~I.} \REVIEW{Phys. Rev.
  A}{95}{2017}{023604}.

\bibitem{std4}
\Name{Bender J., Zohar E., Farace A. \and Cirac J.~I.} \REVIEW{New J.
  Phys.}{20}{2018}{093001}.

\bibitem{std5}
\Name{Kivlichan I.~D., Gidney C., Berry D.~W., Wiebe N., McClean J., Sun W.,
  Jiang Z., Rubin N., Fowler A., Aspuru-Guzik A., Neven H. \and Babbush R.}
  \REVIEW{{Quantum}}{4}{2020}{296}.

\bibitem{timtd1}
\Name{Kim K., Korenblit S., Islam R., Edwards E.~E., Chang M.-S., Noh C.,
  Carmichael H., Lin G.-D., Duan L.-M., Wang C. C.~J., Freericks J.~K. \and
  Monroe C.} \REVIEW{New Journal of Physics}{13}{2011}{105003}.

\bibitem{td3}
\Name{Martinez E.~A. \etal} \REVIEW{Nature}{534}{2016}{516}.

\bibitem{timtd2}
\Name{Gustafson E., Meurice Y. \and Unmuth-Yockey J.} \REVIEW{Phys. Rev.
  D}{99}{2019}{094503}.

\bibitem{d4}
\Name{Lamm H., Lawrence S. \and Yamauchi Y.} \REVIEW{Phys. Rev.
  D}{100}{2019}{034518}.

\bibitem{classicalsimulation1}
\Name{Jones T., Brown A., Bush I. \and Benjamin S.~C.} \REVIEW{Scientific
  Reports}{9}{2019}{10736}.

\bibitem{omelyan}
\Name{Omelyan I.~P., Mryglod I.~M. \and Folk R.} \REVIEW{Phys. Rev.
  E}{65}{2002}{056706}.

\bibitem{fg1}
\Name{Suzuki M.} \REVIEW{Physics Letters A}{201}{1995}{425}.

\bibitem{fg2}
\Name{Omelyan I.~P., Mryglod I.~M. \and Folk R.} \REVIEW{Phys. Rev.
  E}{66}{2002}{026701}.

\bibitem{fg3}
\Name{Kennedy A.~D., Clark M.~A. \and Silva P.~J.}
  \REVIEW{PoSL}{AT2009}{2009}{021}.

\bibitem{gradientascent1}
\Name{Khaneja N., Reiss T., Kehlet C., Schulte-Herbrüggen T. \and Glaser
  S.~J.} \REVIEW{Journal of Magnetic Resonance}{172}{2005}{296}.

\bibitem{gradientascent4}
\Name{Wu R.-B., Chu B., Owens D.~H. \and Rabitz H.} \REVIEW{Phys. Rev.
  A}{97}{2018}{042122}.

\bibitem{gradientascent2}
\Name{Gharibnejad H., Schneider B., Leadingham M. \and Schmale H.}
  \REVIEW{Computer Physics Communications}{252}{2020}{106808}.

\bibitem{gradientascent3}
\Name{Raza A., Hong C., Wang X., Kumar A., Shelton C.~R. \and Wong B.~M.}
  \REVIEW{Computer Physics Communications}{258}{2021}{107541}.

\bibitem{optimization1}
\Name{Childs A.~M., Ostrander A. \and Su Y.} \REVIEW{{Quantum}}{3}{2019}{182}.

\bibitem{optimization2}
\Name{Hadfield S. \and Papageorgiou A.} \REVIEW{New Journal of
  Physics}{20}{2018}{043003}.

\bibitem{optimization3}
\Name{Ouyang Y., White D.~R. \and Campbell E.~T.}
  \REVIEW{{Quantum}}{4}{2020}{235}.

\bibitem{nested}
\Name{Shcherbakov D., Ehrhardt M., Finkenrath J., G\"unther M., Knechtli F.
  \and Peardon M.} \REVIEW{Commun. Comput. Phys.}{21}{2017}{1141}.

\bibitem{threeterm}
\Name{Barthel T. \and Zhang Y.} \REVIEW{Annals of Physics}{418}{2020}{168165}.

\bibitem{KogutReview}
\Name{Kogut J.~B.} \REVIEW{Rev. Mod. Phys.}{51}{1979}{659}.

\bibitem{TIMBook}
\Name{Sei~Suzuki, Jun-ichi~Inoue B. K. C.~a.} \Book{Quantum Ising Phases and
  Transitions in Transverse Ising Models} 2nd Edition Lecture Notes in Physics
  862 (Springer-Verlag Berlin Heidelberg) 2013.

\bibitem{timkc1}
\Name{Rieger H. \and Kawashima N.} \REVIEW{Euro. Phys. B}{9}{1999}{233}.

\bibitem{timkc2}
\Name{Hamer C.~J.} \REVIEW{J. Phys. A}{33}{2000}{6683}.

\bibitem{timkc3}
\Name{Bl\"ote H. W.~J. \and Deng Y.} \REVIEW{Phys. Rev. E}{66}{2002}{066110}.

\bibitem{timkc4}
\Name{Evenbly G. \and Vidal G.} \REVIEW{Phys. Rev. Lett.}{102}{2009}{180406}.

\bibitem{z2lgt}
\Name{Wegner F.~J.} \REVIEW{J. Math. Phys.}{12}{1971}{2259}.

\bibitem{Fradkin1978}
\Name{Fradkin E.~H. \and Susskind L.} \REVIEW{Phys. Rev. D}{17}{1978}{2637}.

\bibitem{quantumlinkmodel}
\Name{Chandrasekharan S. \and Wiese U.-J.} \REVIEW{Nuclear Physics
  B}{492}{1997}{455}.

\bibitem{z2dualtoTIM}
\Name{Balian R., Drouffe J.~M. \and Itzykson C.} \REVIEW{Phys. Rev.
  D}{11}{1975}{2098}.

\bibitem{z2Sachdev}
\Name{Sachdev S.} \REVIEW{Rept. Prog. Phys.}{82}{2019}{014001}.

\bibitem{z2quantum}
\Name{Cui X., Shi Y. \and Yang J.-C.} \REVIEW{JHEP}{08}{2020}{160}.

\end{thebibliography}
\bibliographystyle{eplbib}

\end{document}